# Underwater metamaterial absorber with impedance-matched composite

**Authors**
Sichao Qu[1†], Nan Gao[1†], Alain Tinel[2], Bruno Morvan[2], Vicente Romero-García[3], Jean-Philippe Groby[3], Ping Sheng[1*]

**Affiliations**
[1]Department of Physics, The Hong Kong University of Science and Technology, Clear Water Bay, Kowloon, Hong Kong, China

[2]Laboratoire Ondes et Milieux Complexes UMR CNRS 6294, UNILEHAVRE, Normandie University, 75 Rue Bellot, 76600 Le Havre, France

[3]Laboratoire d'Acoustique de l'Université du Mans (LAUM), UMR 6613, Institut d'Acoustique - Graduate School (IA-GS), CNRS, Le Mans Université, France

†These authors contributed equally to this work
*Email: sheng@ust.hk

**Abstract**
By using a structured tungsten-polyurethane composite that is impedance-matched to water while simultaneously having a much slower longitudinal sound speed, we have theoretically designed, and experimentally realized, an underwater acoustic absorber exhibiting high absorption from 4 to 20 kHz, measured in a 5.6 m×3.6 m water pool with the time-domain approach. The broadband functionality is achieved by optimally engineering the distribution of the Fabry-Perot resonances, based on an integration scheme, to attain impedance matching over a broad frequency range. The average thickness of the integrated absorber, 8.9 mm, is in the deep subwavelength regime (~$\lambda/42$ at 4 kHz) and close to the causal minimum thickness of 8.2 mm that is evaluated from the simulated absorption spectrum. The structured composite represents a new type of acoustic metamaterials that has high acoustic energy density and promises broad underwater applications.

**Teaser**
This work reports a very effective, broadband underwater acoustic absorber that is thin and impedance-matched to water.

# MAIN TEXT

## Introduction

Underwater acoustics represents an area of study that is important for the subsurface exploration and object-imaging in rivers and oceans that occupy the majority of Earth's surface. The efficient absorption of low-frequency underwater acoustic waves has especially attracted strong interest, notably for the applications in underwater sensing and stealth technologies (*1-3*). In spite of its apparent importance, however, this topic is nowhere as intensely pursued as either airborne audible sound (*4-7*) or ultrasound (*8, 9*). The latter represents the higher frequency branch of the waterborne acoustic waves that has found widespread use in medical applications. One reason for this state of affairs is the difficulty of experimental measurements, owing to the large wavelength involved and the required large impedance mis-match of the solid material for a water impedance tube (*10, 11*). As a result, considerable studies on underwater absorption are only limited to theoretical analyses and numerical calculations (*12-20*), whose idealized assumptions may not hold in practical scenarios; while almost all the existing experimental works (*21-28*), measured in the water impedance tube, are based on small samples, which may not reflect the true performance in complex environments (*29*). There is simply a lack of research works based on large-scale samples, measured in water pools. Besides the experimental obstacles, designing an underwater absorber itself can be a challenge for theoretical modelling, due to the diversity of solid elastic vibration modes (*30*) that can give rise to difficulty in focusing on the absorption functionality absent of any undesired features. In addition, the acoustic energy density of the conventional materials (*31, 32*) for underwater applications is relatively low, which hinders the efficient dissipation of the low-frequency waves within an acoustically-thin sample. In other words, the potential of reducing the thickness of the underwater absorber has not yet been fully explored.

In this work, we present a novel metamaterial absorber with structured impedance-matched composite, which can offer a solution to the aforementioned challenges in underwater absorption. Our composite comprises tungsten particles dispersed in a polyurethane (PU) polymer matrix, geometrically structured into slender rods, thereby enabling to treat them as one-dimensional solids. As the longitudinal speed in a rod is governed by its Young's modulus $E_c$, for the PU matrix it can be easily tuned to be lower than the bulk modulus of water $B_w$. Meanwhile, the dispersed tungsten particles can contribute to a considerably larger mass density $\rho_c$ than that of water, $\rho_w$. It follows that the characteristic impedance of our structured composite can be easily tuned to match that of water. Simultaneously, such a composite material would have a much slower longitudinal wave speed than that in water, thereby implying a larger acoustic density of states at low frequencies. As a result, when the rods are backed by a hard reflecting boundary, the Fabry-Perot (FP) resonances can be realized with a much thinner thickness than that of the conventional materials.

In what follows, we delineate the process of synthesizing tungsten-polyurethane composite to attain the desired properties. To achieve broadband impedance matching, we employ an integration scheme of multiple FP resonances and experimentally verify the absorption performance of a 0.92 m×0.92 m sample in a water pool. It will be shown that good agreement between the simulation and experiment is obtained, with an average absorption over 90% from 4 to 20 kHz and an average sample thickness of only 8.9 mm, which is close to the minimum sample thickness as dictated by the causality constraint (*5*).

# Results

## The impedance-matching composite and slow sound effect

We introduce a dimensionless scaling factor $\alpha$ ($\alpha > 1$) to simultaneously tune the following parameters of the composite rod—the density $\rho_c$ and the longitudinal modulus $M$:

$$\rho_c \to \alpha \rho_w, \quad M \to \frac{B_w}{\alpha}, \tag{1}$$

where the water density $\rho_w = 1$ g/cm$^3$ and bulk modulus $B_w = 2.16 \times 10^9$ Pa. In this manner, the composite's characteristic acoustic impedance $\sqrt{\rho_c M}$ is matched with that of water $\sqrt{\rho_w B_w}$. Also, the sound speed in the composite will be slower, i.e., $v_c = \sqrt{\frac{M}{\rho_c}} = \frac{1}{\alpha}\sqrt{\frac{B_w}{\rho_w}} = \frac{v_w}{\alpha}$, while the wavelength is compressed as well, i.e., $\lambda = \frac{v_w}{\alpha f} = \frac{\lambda_w}{\alpha}$. In fact, we treat the slender composite rod as one-dimensional solid materials since its length $L$ is longer than its lateral size $a = 5$ mm. Therefore, the longitudinal modulus $M$ can be approximated as Young's modulus $E_c$ of the composite rod with the free boundary condition on the four sidewalls of the rod and rigid backing condition at the bottom [Fig. 1(a)]. This can be evidenced by the simulation based on Finite-Element-Method (FEM), where we evaluate the displacement of the upper surface of the rod $\Delta L$, under a static pressure modulation $\Delta p$. We directly set the Young's modulus $E_c = B_w/\alpha$ and compare it with the retrieved effective longitudinal modulus, given by $M = L\Delta p/\Delta L$. The simulation results, displayed in Fig. 1(b), confirm the one-dimensional model assumption, because $M \cong E_c$ for all values of $\alpha$.

If we launch a harmonic plane wave onto the composite rod [Fig. 1(a)], the FP resonance (5) will occur when a quarter wavelength is equal to the rod length $L$. Here, by fixing $f$ to be the target frequency, the length of the FP rod $L$ is proportional to the sound speed in composite materials, with $L = \frac{v_c}{4f} = \frac{v_w}{4\alpha f} = \frac{L_0}{\alpha}$. This indicates that by using slow sound materials ($\alpha > 1$), we can reduce the FP resonator's length without changing the resonant frequency. Furthermore, we investigate the surface response spectrum as characterized by the Green function (6, 33), defined by: averaged displacement $x$/pressure modulation $p$, at the top surface of the rod. As an example, we choose the target frequency $f = 3.5$ kHz, with $L_0 = 0.1$ m. It is seen that when the parameters are scaled as in Eq. (1) with $\alpha = 1$ or 5.5, the relevant Green function, $G(\omega)$, plotted as a function of frequency in Fig. 1(c-d), remains exactly the same for the two cases as evidenced by the complete overlap of the blue continuous and red dashed lines. In other words, at the theoretical level, there exists an opportunity that the same surface response can be supported with a much shorter length, $L = L_0/\alpha$, when $\alpha$ is large. In the following section, we show such material properties with $\alpha = 5.5$ given in Eq. (1) can be realized experimentally.

## Tuning the composite's acoustic properties

We target a composite with high acoustic energy density as specified by $\alpha = 5.5$. In the literature, an impedance-matched material is usually denoted as 'Rho-C rubber' (1, 34, 35), which usually means that both the material's density and sound speed are close to those of water. However, for our structured composite rods, while the effective impedance can be matched to that of water, the wave speed is slower and the acoustic energy density (36) is $\alpha$ times higher, given by

$$\varepsilon = \frac{1}{2}\rho_c v^2 + \frac{1}{2}\frac{1}{M}p^2 = \alpha\left(\frac{1}{2}\rho_w v^2 + \frac{1}{2}\frac{1}{B_w}p^2\right), \quad (2)$$

where $p$ and $v$ denote the acoustic pressure modulation and the particle displacement velocity, respectively, and the term in the bracket is the energy density in water or the Rho-C rubber.

The ingredients for synthesizing the composite are tungsten granules (W), uniformly dispersed in polyurethane polymer (PU) [Fig. 2(a)]. Here the PU polymer was composed of a softer PU rubber and a harder PU resin [Fig. 2(b)]. The tungsten, with its density of 19.3 g/cm³, was used for the purpose of creating a composite with large density, while the PU polymer was chosen for low and tunable modulus. The total mass of the composite is the summation of all the ingredients:

$$\begin{cases} M_{\text{total}} = M_{\text{PU}} + M_{\text{W}} = (R_1 + 1)M_{\text{W}}, \\ M_{\text{PU}} = M_{\text{resin}} + M_{\text{rubber}} = (R_2 + 1)M_{\text{rubber}}, \end{cases} \quad (3)$$

where $R_1 = M_{\text{PU}}/M_{\text{W}}$, $R_2 = M_{\text{resin}}/M_{\text{rubber}}$, and the subscript 'W' denotes tungsten.

We show that by tuning $R_1$ and $R_2$ in two steps, the target density and longitudinal modulus can be attained. We first mixed tungsten granules with PU resin and PU rubber separately ($\rho_{\text{resin}} = \rho_{\text{rubber}} \cong 1$ g/cm³). Since the density of PU rubber is the same as PU resin, the outcome of the mixed gels coincides when the ratio between the PU polymer and tungsten is the same (Fig. 2(c)). By adjusting the ratio of tungsten to be $R_1 = 0.0145$, we obtained $\rho_c = 5.5$ g/cm³ $= 5.5\rho_w$. Next, by fine tuning the ratio $R_2$, we manipulated the Young's modulus of the composite without changing the density. It turned out that if $R_2 = 0.524$ [Fig. 2(d)], the resulting modulus $E_c = 0.37 \times 10^9$ Pa $\cong B_w/5.5$ [Fig. 2(e)]. In this manner, we realized the composite with a high mass density and low bulk modulus, while keeping the impedance matched to that of water, i.e., $\sqrt{\rho_c E_c} = \sqrt{\rho_w B_w}$. Theoretically, the target $\alpha$ can be higher than 5.5, as long as we are still working below the percolation threshold (33) of the tungsten granules. However, in order to ensure the uniformity of the composite sample, $\alpha = 5.5$ was fixed for the subsequent theoretical design and sample fabrication.

**FP resonators and the integration scheme for broadband absorption**

A FP resonator is simply a slender solid rod with a rigid reflecting boundary at the bottom end. The resonance condition is attained when one-quarter of the relevant wavelength matches the length of the rod. Since the relevant wavelength in the composite rod is much shorter, it follows that the length of the resonators can also be much shorter, thereby leading to a thinner absorber. The lateral size of the resonators is very small relative to the wavelength (i.e., $a = 5$ mm $\ll \lambda$). Therefore, only the longitudinal modes should be taken into consideration and the surface impedance of the individual $n$th FP resonator in an integrated array has the simple form $Z_n = iZ_w \cot(\omega L_n/v_c)$. Note that we use the harmonic time dependence of $\exp(-i\omega t)$ throughout our manuscript. The Green function (6, 33), defined as the surface response of $n$th composite rod, is given by $G_n = u_n/(-i\omega p) = 1/(-i\omega Z_n)$, where $u_n$ denotes the displacement velocity on the upper surface. With some mathematical manipulation (37) (see details in SI Appendix, Supporting Text 1), $G_n$ may be expressed in the Lorentzian form

$$G_n(\omega) \cong \frac{\chi_n}{\Omega_n^2 - \omega^2 - i\beta\omega}, \quad (4)$$

where the oscillation strength $\chi_n = 2/(\rho_c L_n) = 2/(\rho_w L_0)$, the resonant angular frequency $\Omega_n = 2\pi f_n = \pi v_c/(2L_n)$, and the factor $\beta$ is introduced to model the dissipation (details are available in *SI Appendix*, Supporting Text 2). We see that Eq. (4) is independent of $\alpha$, which is consistent with the results shown in Fig. 1(c-d).

To realize broadband impedance matching, we integrate FP resonators with nine different lengths $(L_1, L_2, ..., L_9)$, as a unit cell [see schematic illustration of the resonator integration in Fig. 3(a-b)]. Earlier works (*5, 38*) have shown that if we adopt the approximation of ignoring the higher-order FP resonances in each of the resonators, then the optimal choice of the resonance frequencies $(f_1, f_2, ..., f_9)$, for yielding a flat, near-total absorption curve is given by the formula $f_n = f_1 e^{2(n-1)/9}$ as plotted in dashed line in Fig. 3(c), with the corresponding $L_n = v_c/(4f_n)$. Here, $L_1 = L_0/5.5$ and $f_1 = 3.5$ kHz. An integration scheme, including the correction due to the higher-order FP resonances in each resonator, is given in the *SI Appendix*, Supporting Text 3, which yields the resonance distribution slightly different from the first-order power law, as shown by the solid line in Fig. 3(c). The final discretized resonances are indicated by the circles in Fig. 3(c) and arrows in Fig. 3(e), leading to the final length recipe to be $L_1 = 18.9$ mm, $L_2 = 15.2$ mm, $L_3 = 12.2$ mm, $L_4 = 9.8$ mm, $L_5 = 7.9$ mm, $L_6 = 6.4$ mm, $L_7 = 4.6$ mm, $L_8 = 3.1$ mm, and $L_9 = 1.7$ mm. It should be noted that while the length is smaller than the lateral size ($a = 5$ mm) for resonators 7-9, simulations have shown the desired scaling characteristics can still hold.

To verify the effect of the integration scheme, we first calculate the averaged surface response of the integrated resonators placed in parallel such that $G_s = \sum_{n=1}^{9} G_n/9$, where $G_n$ is given by Eq. (4). Here the factor $1/9$ denotes the area fraction occupied by each resonator surface facing the incident wavefront. The overall surface impedance of the sample array is given by:

$$Z_s(\omega) = \frac{1}{-i\omega G_s} = \left(\frac{1}{9}\sum_{n=1}^{9}\frac{1}{Z_n}\right)^{-1}, \qquad (5)$$

which is shown by the solid curve in Fig. 3(d). In Eq. (5), as an approximation, we have ignored the coupling effects between the adjacent rods (*5, 6*), through the evanescent waves. It is seen that broadband impedance matching is achieved with the chosen rod lengths. The real part of the impedance is close to $Z_w$ beyond 3.5 kHz, and the imaginary part vanishes as the frequency increases. The absorption coefficient can be obtained by inserting Eq. (5) into the following expression:

$$A(\omega) = 1 - \left|\frac{Z_s - Z_w}{Z_s + Z_w}\right|^2, \qquad (6)$$

where the water impedance $Z_w = \sqrt{\rho_w B_w}$. In Fig. 3(e), we show the resulting broadband absorption, given by Eq. (6), which starts around 4 kHz (slightly higher than $f_1$). We also compare the theoretical prediction with that of full-wave FEM simulations, plotted as circles in Fig. 3(e). Good agreement between theory predictions and simulations is seen. It should be noted that two neighboring rods are separated by a ~50 micron air gap [Fig. 3(b)], with the soft water-repellent glue seal on top. The air gap is essentially the average separation between two macroscopically flat surfaces of the two touching objects, owing to the usual amount of surface asperities and microscopic undulations. While there can be multiple contact points of the two surfaces, longitudinal vibration modes in the nearby rods are essentially decoupled from each other. In Fig. 3(f), we plot the velocity fields inside the FP resonators from $f_1$ to $f_6$. It is not surprising that at the resonance frequency $f_n$, the rod with $L_n$ displays the response with largest surface velocity. For the frequencies in-between the FP resonances (or higher frequencies beyond

$f_6$), the impedance matching condition is realized as well, but more than one resonator can be excited.

## The causality constraint on sample thickness

Due to the fundamental causal nature (*39-41*) of an absorber's response to the incident wave, there is a minimum sample thickness associated with any given absorption spectrum $A(\lambda)$. For acoustic systems, this causal constraint takes the form of an inequality (*5, 6*):

$$\bar{d} \geq d_{\min} = \frac{1}{4\pi^2} \frac{B_{\text{eff}}}{B_0} \int_0^\infty \left|\ln\left(1 - A(\lambda)\right)\right| d\lambda, \tag{7}$$

where $\bar{d}$ denotes the average thickness of the absorber, and at the long wavelength limit, $B_{\text{eff}}/B_w \cong E_c/B_w = 1/5.5$ in the present case. From Eq. (7), we again see that by using the slow sound medium, the lower bound thickness, $d_{\min}$, can be significantly reduced, which is consistent with the results shown in Fig. 1(c-d). We can insert the simulated absorption in Fig. 3(e) into Eq. (7) to obtain $d_{\min} = 8.2$ mm, which is close to the average thickness ($\bar{d} = \sum_{n=1}^{9} L_n/9 = 8.9$ mm). Hence, at the stage of theoretical design, we conclude that the performance of our absorber has approached the causality limit. Although the causally optimal broadband absorbers (COBA) have been realized in airborne acoustic (*5, 6*) and electromagnetic systems (*41*), we demonstrate here that the causal lower limit can be modified, or more specifically lowered, by using impedance-matched composite with slow-sound properties.

## Sample fabrication considerations

As for sample fabrication, we have structured the composite into rod shapes with pre-designed lengths by using molds [Fig. 4(a)]. The fabrication procedures of the rods are given in the *SI Appendix*, Supporting Text 4. Since the FP resonators require a reflecting backing substrate, we have also fabricated a stainless-steel base, with stepped stairs that is complementary to the lengths of the rods, so that the upper surface of the sample is flat [Fig. 4(a-b)]. The total sample thickness, including the stainless-steel base, was designed to be ~0.069 m, so as to eliminate transmission beyond 3.5 kHz. We let neighboring rods come into contact naturally, so that there must exist some small amount of air space between the rods. We brushed a thin layer of the hydrophobic soft glue to trap the air and to prevent water from seeping into the gaps [Fig. 4(b)]. Due to the heavy mass of the stainless-steel base, we divided the sample into 3 × 3 units, so as to facilitate easy assembly into the whole with a dimension of 0.92 m × 0.92 m [Fig. 4(b-c)].

## Experimental verification in a water pool

To test the absorption of the sample in an underwater environment, we have carried out the measurements in a water pool with the configuration shown in Fig. 4(d). The sample, source, and hydrophone were placed in the middle of the 5.6 m×3.6 m water pool to minimize the interference of the boundaries. A burst signal was emitted by the directional source, reflected by the sample and received by a hydrophone at two different positions for collecting the reflection and transmission data, plotted in Fig. 4(e) by circles. To compare with the experimental data, we take into account the stainless-steel base in the simulations (instead of the rigid backing condition), so as to allow some transmission of the incident wave, which is plotted as green dashed lines in Fig. 4(e).

In the low frequency regime below 0.5 kHz, the transmission dominates due to the by-passing effect of the long wavelength. With increasing frequency, reflection appears and reaches its maximum around 2 kHz. If the sample is made of pure stainless steel, the reflection will keep increasing and reach almost unity at higher frequencies, but due to the meta-structures on the front side of the sample, the strong reflection turns into the absorption inside the high-energy-density composite. We obtain an averaged absorption (from 4 to 20 kHz) of 97.6% for the simulation and 90.3% for the experiment [Fig. 4(e)]. It should be noted that the averaged sample thickness is only 8.9 mm, which is only 1/42 wavelength at 4 kHz, while for the longest rod, the same value is 1/20.

## Discussion

By comparing our work with the traditional absorbers, such as the Alberich coatings (*12*) and penta-mode structures (*18*), we find that our design scheme is straightforward and more effective because only the longitudinal modes are involved, achieved by both the rod geometry separated by air gaps, and the subwavelength lateral dimension of each rod. In this manner, the lateral modes can only exist in the form of evanescent waves, which cannot couple to the incident wave (*42*). In fact, further simulation shows that smaller lateral size of the rods not only helps improving the higher frequency absorption but also enhances the performance under oblique incidence (*43*) (results are available in *SI Appendix,* Supporting Text 5). Unlike the air-bubble-based absorber (*12, 21, 44, 45*), we do not rely on the resonant modes shaped by air cavities to achieve absorption. Therefore, the performance will not significantly suffer from the hydrostatic pressure (see more simulation results in *SI Appendix,* Supporting Text 6). It should be noted that our composite's intrinsic properties ensure that the slow sound effect holds for the whole frequency spectrum, which differs from the resonance-based slow sound absorber (*46-48*), for which the enhanced absorption occurs near the resonant frequencies. As for the absorbing bands, the starting frequency is actually tunable, and can be easily shifted down to the lower sub-kilohertz regime. This can be done by multiplying the current rod lengths by a factor larger than 1 (see absorption results in *SI Appendix,* Supporting Text 7). In our experiment, we just measured from 3 kHz to 20 kHz, due to the limitations of the pool size and the performance of the sound sources. However, through simulations, we do not see any reason why our absorber cannot extend to lower and higher frequencies.

In conclusion, our work not only offers an integration scheme for broadband impedance-matching, but also proposes a novel type of impedance-matched and high-energy-density composite that can realize excellent underwater absorption with thin sample thickness. The former provides a design recipe for solving the narrow band problem of resonance-based underwater acoustic absorbers, while the latter makes a pioneering step in reducing the thickness of the absorber without losing the broadband absorption performance. The absorption performance of a large-size sample was verified by measurements in a water pool, with an averaged absorption of 90.3% from 4 kHz to at least 20 kHz. The materials and the design methodology adopted in this work may provide a new route in designing versatile devices with broadband features, such as camouflage stealth materials (*2, 3, 49*), tunable waveguides (*50, 51*), metamaterial-based lens (*8, 52, 53*), impedance transformer (*54, 55*), etc.

## Materials and Methods

## Sample fabrication and characterization

The composite materials were fabricated by using the ingredients described in the main text. The Young's modulus of the polyurethane resin (model Task 6) is $1 \times 10^9$ Pa and that of the polyurethane rubber (model Simpact 85A) is much softer with a value of $6.83 \times 10^6$ Pa. The process of mould-making and demoulding can be found in *SI Appendix*, Supporting Text 4. The composite sample was fabricated into the long strip geometry [see Fig. 2(b)] for the tensile test. The tensile equipment we used is Electronic Universal Testing Machine WDW-10 from Jinan Jinyinfeng Instrument Company. The stainless-steel base of the large-scale sample (for the purpose of achieving longitudinal wave reflection) was precisely fabricated using the Wire Electrical Discharge Machining (WEDM). The averaged thickness of the base with the stepped surfaces is 0.06 m, so that the transmitted wave beyond 4 kHz can be significantly suppressed. The density of the stainless steel is 7.5 g/cm$^3$. By combining the mass of composite rods with the base, the total mass of the sample is around 420 kg. The Young's modulus of the soft glue is in the order of $10^5$ Pa, with a negligible shear modulus.

## Simulation methods and setups

In numerical simulations, the commercial software COMSOL Multiphysics was adopted for efficient computation. The static pressure test in Fig. 1(a) was simulated by the Solid Mechanics module. For the simulations involving the evaluation of absorption, the Solid Mechanics module was combined with the Pressure Acoustics module. The former was responsible for the domain of composite rods, the latter was responsible for the water domain. For the solid-fluid interfaces, continuity boundary conditions were applied for the displacement and stress fields. The lateral boundaries of the simulation domain were set to be the Floquet periodic boundary condition. The four side walls of the composite rods were applied with free boundary condition for the simulation whose results are shown in Fig. 3(d-e), as an approximation. However, for Fig. 4(e), the air gaps were occupied by the acoustic domain and the free boundaries were replaced by the solid-air interface. Also, the air gaps and the water are gapped by a thin layer of soft glue with Young's modulus to be $\sim 10^5$ Pa. Meanwhile, the stainless-steel base and the water domain at the back side were also added, for collecting the transmission information. The mesh size was set to be smaller than 1/6 of a wavelength in order to ensure accuracy.

## Water pool measurements

We adopted time-domain approach to measure the transmission and reflection coefficients in a water pool. A pistonic Piezoelectric Underwater emitter (Lubell model 9162T) was used for wave generation while an omnidirectional hydrophone (Brüel & Kjaer Type 8105), associated with a signal amplifier conditioner (Nexus Type 2692), was used as receiver. A signal burst windowed by a Hanning function was sent to the source. Each group of the measured signals was averaged 300 times to improve the signal-to-noise ratio. In order to handle the parasitic echoes coming from multiple reflections on the walls of the pool, we used the following procedures: measurements were done in the water tank at two positions P1 (front side) and P2 (back side). For each frequency, we carried out the two groups of measurements (i.e., with and without the sample) and then collected the data at P1 and P2. The data without the sample act as the reference signals $S_{\text{ref}-\text{P1}}$ (including the incident wave and spurious signals from the reflections by pool walls) and $S_{\text{ref}-\text{P2}}$ (approximately treated as the incident signal strength). With the sample present, the signals at P1 and P2 are denoted by $S_{\text{P1}}$ and $S_{\text{P2}}$, respectively. Therefore, after projecting the signals into the frequency domain by Fourier transform, we can calculate the reflectance $R = |(S_{\text{P1}} - S_{\text{ref}-\text{P1}})/S_{\text{ref}-\text{P2}}|^2$ and the transmittance $T = |S_{\text{P2}}/S_{\text{ref}-\text{P2}}|^2$.

## References

1. D. Lee *et al.*, Underwater stealth metasurfaces composed of split-orifice–conduit hybrid resonators. *Journal of Applied Physics* **129**, 105103 (2021).
2. G. Yu, Y. Qiu, Y. Li, X. Wang, N. Wang, Underwater Acoustic Stealth by a Broadband 2-Bit Coding Metasurface. *Physical Review Applied* **15**, 064064 (2021).
3. Y. Gao *et al.*, Hydrogel microphones for stealthy underwater listening. *Nature communications* **7**, 1-11 (2016).
4. N. Jiménez, V. Romero-García, V. Pagneux, J.-P. Groby, Rainbow-trapping absorbers: Broadband, perfect and asymmetric sound absorption by subwavelength panels for transmission problems. *Scientific reports* **7**, 1-12 (2017).

5. M. Yang, S. Chen, C. Fu, P. Sheng, Optimal sound-absorbing structures. *Materials Horizons* **4**, 673-680 (2017).
6. M. Yang, P. Sheng, Sound absorption structures: From porous media to acoustic metamaterials. *Annual Review of Materials Research* **47**, 83-114 (2017).
7. J. Li, W. Wang, Y. Xie, B.-I. Popa, S. A. Cummer, A sound absorbing metasurface with coupled resonators. *Applied Physics Letters* **109**, 091908 (2016).
8. S. Zhang, C. Xia, N. Fang, Broadband acoustic cloak for ultrasound waves. *Physical review letters* **106**, 024301 (2011).
9. Y. Jin, R. Kumar, O. Poncelet, O. Mondain-Monval, T. Brunet, Flat acoustics with soft gradient-index metasurfaces. *Nature communications* **10**, 1-6 (2019).
10. P. S. Wilson, R. A. Roy, W. M. Carey, An improved water-filled impedance tube. *The Journal of the Acoustical Society of America* **113**, 3245-3252 (2003).
11. M. Oblak, M. Pirnat, M. Boltežar, An impedance tube submerged in a liquid for the low-frequency transmission-loss measurement of a porous material. *Applied Acoustics* **139**, 203-212 (2018).
12. D. Zhao, H. Zhao, H. Yang, J. Wen, Optimization and mechanism of acoustic absorption of Alberich coatings on a steel plate in water. *Applied Acoustics* **140**, 183-187 (2018).
13. J. Wen et al., Effects of locally resonant modes on underwater sound absorption in viscoelastic materials. *The Journal of the Acoustical Society of America* **130**, 1201-1208 (2011).
14. Y. Zhang, J. Pan, K. Chen, J. Zhong, Subwavelength and quasi-perfect underwater sound absorber for multiple and broad frequency bands. *The Journal of the Acoustical Society of America* **144**, 648-659 (2018).
15. C. Yu, M. Duan, W. He, F. Xin, T. Lu, Underwater anechoic layer with parallel metallic plate insertions: theoretical modelling. *Journal of Micromechanics and Microengineering*, (2021).
16. L. B. Wang, C. Z. Ma, J. H. Wu, A thin meta-structure with multi-order resonance for underwater broadband sound absorption in low frequency. *Applied Acoustics* **179**, 108025 (2021).
17. H. Wu, H. Zhang, C. Hao, Reconfigurable spiral underwater sound-absorbing metasurfaces. *Extreme Mechanics Letters* **47**, 101361 (2021).
18. Y. Gu, H. Long, Y. Cheng, M. Deng, X. Liu, Ultrathin Composite Metasurface for Absorbing Subkilohertz Low-Frequency Underwater Sound. *Physical Review Applied* **16**, 014021 (2021).
19. J. Zhong et al., Theoretical requirements and inverse design for broadband perfect absorption of low-frequency waterborne sound by ultrathin metasurface. *Scientific reports* **9**, 1-11 (2019).
20. M. Duan, C. Yu, F. Xin, T. J. Lu, Tunable underwater acoustic metamaterials via quasi-Helmholtz resonance: From low-frequency to ultra-broadband. *Applied Physics Letters* **118**, 071904 (2021).
21. Y. Fu, J. Fischer, K. Pan, G. H. Yeoh, Z. Peng, Underwater sound absorption properties of polydimethylsiloxane/carbon nanotube composites with steel plate backing. *Applied Acoustics* **171**, 107668 (2021).
22. H. Jiang, Y. Wang, Phononic glass: A robust acoustic-absorption material. *The Journal of the Acoustical Society of America* **132**, 694-699 (2012).
23. Y. Gu, H. Zhong, B. Bao, Q. Wang, J. Wu, Experimental investigation of underwater locally multi-resonant metamaterials under high hydrostatic pressure for low frequency sound absorption. *Applied Acoustics* **172**, 107605 (2021).
24. B. Yuan, W. Jiang, H. Jiang, M. Chen, Y. Liu, Underwater acoustic properties of graphene nanoplatelet-modified rubber. *Journal of Reinforced Plastics and Composites* **37**, 609-616 (2018).
25. J. Heng et al., A wide band strong acoustic absorption in a locally network anechoic coating. *Chinese Physics Letters* **26**, 106202 (2009).
26. H. Jiang et al., Locally resonant phononic woodpile: A wide band anomalous underwater acoustic absorbing material. *Applied Physics Letters* **95**, 104101 (2009).
27. H. Meng, J. Wen, H. Zhao, X. Wen, Optimization of locally resonant acoustic metamaterials on underwater sound absorption characteristics. *Journal of Sound and Vibration* **331**, 4406-4416 (2012).
28. L. Li, Z. Zhang, Q. Huang, S. Li, A sandwich anechoic coating embedded with a micro-perforated panel in high-viscosity condition for underwater sound absorption. *Composite Structures* **235**, 111761 (2020).
29. Y. Zhang, K. a. Chen, X. Hao, Y. Cheng, A review of underwater acoustic metamaterials. *Chinese Science Bulletin* **65**, 1396-1410 (2020).

30. G. Ma *et al.*, Polarization bandgaps and fluid-like elasticity in fully solid elastic metamaterials. *Nature communications* **7**, 1-8 (2016).
31. Y. Fu, I. I. Kabir, G. H. Yeoh, Z. Peng, A review on polymer-based materials for underwater sound absorption. *Polymer Testing*, 107115 (2021).
32. V. Jayakumari, R. Shamsudeen, R. Rajeswari, T. Mukundan, Viscoelastic and acoustic characterization of polyurethane‐based acoustic absorber panels for underwater applications. *Journal of Applied Polymer Science* **136**, 47165 (2019).
33. P. Sheng, *Introduction to wave scattering, localization and mesoscopic phenomena*. (Springer Science & Business Media, 2006), vol. 88.
34. L. Adair, R. Cook, Acoustic Properties of Rho‐C Rubber and ABS in the Frequency Range 100‐kHz‐2 MHz. *The Journal of the Acoustical Society of America* **54**, 1763-1765 (1973).
35. R.-M. Guillermic, M. Lanoy, A. Strybulevych, J. H. Page, A PDMS-based broadband acoustic impedance matched material for underwater applications. *Ultrasonics* **94**, 152-157 (2019).
36. S. Qu, P. Sheng, Minimizing Indoor Sound Energy with Tunable Metamaterial Surfaces. *Physical Review Applied* **14**, 034060 (2020).
37. T. Gamelin, *Complex analysis*. (Springer Science & Business Media, 2003).
38. M. Yang, P. Sheng, An integration strategy for acoustic metamaterials to achieve absorption by design. *Applied Sciences* **8**, 1247 (2018).
39. J. D. Jackson, Classical electrodynamics. *American Institute of Physics* **15**, 62-62 (2009).
40. K. N. Rozanov, Ultimate thickness to bandwidth ratio of radar absorbers. *IEEE Transactions on Antennas and Propagation* **48**, 1230-1234 (2000).
41. S. Qu, Y. Hou, P. Sheng, Conceptual-based design of an ultrabroadband microwave metamaterial absorber. *Proceedings of the National Academy of Sciences* **118**,  (2021).
42. G. Ma, M. Yang, S. Xiao, Z. Yang, P. Sheng, Acoustic metasurface with hybrid resonances. *Nature materials* **13**, 873-878 (2014).
43. N. Jiménez, W. Huang, V. Romero-García, V. Pagneux, J.-P. Groby, Ultra-thin metamaterial for perfect and quasi-omnidirectional sound absorption. *Applied Physics Letters* **109**, 121902 (2016).
44. A. Bretagne, A. Tourin, V. Leroy, Enhanced and reduced transmission of acoustic waves with bubble meta-screens. *Applied Physics Letters* **99**, 221906 (2011).
45. V. Leroy *et al.*, Superabsorption of acoustic waves with bubble metascreens. *Physical Review B* **91**, 020301 (2015).
46. J.-P. Groby, R. Pommier, Y. Aurégan, Use of slow sound to design perfect and broadband passive sound absorbing materials. *The Journal of the Acoustical Society of America* **139**, 1660-1671 (2016).
47. J.-P. Groby, W. Huang, A. Lardeau, Y. Aurégan, The use of slow waves to design simple sound absorbing materials. *Journal of Applied Physics* **117**, 124903 (2015).
48. N. Jiménez, V. Romero-García, V. Pagneux, J.-P. Groby, Quasiperfect absorption by subwavelength acoustic panels in transmission using accumulation of resonances due to slow sound. *Physical Review B* **95**, 014205 (2017).
49. H. Yuk *et al.*, Hydraulic hydrogel actuators and robots optically and sonically camouflaged in water. *Nature communications* **8**, 1-12 (2017).
50. N. Gao, S. Qu, L. Si, J. Wang, W. Chen, Broadband topological valley transport of elastic wave in reconfigurable phononic crystal plate. *Applied Physics Letters* **118**, 063502 (2021).
51. N. Gao, J. Li, R. Bao, W. Chen, Harnessing uniaxial tension to tune Poisson's ratio and wave propagation in soft porous phononic crystals: an experimental study. *Soft Matter* **15**, 2921-2927 (2019).
52. X. Su, A. N. Norris, C. W. Cushing, M. R. Haberman, P. S. Wilson, Broadband focusing of underwater sound using a transparent pentamode lens. *The Journal of the Acoustical Society of America* **141**, 4408-4417 (2017).
53. A. Allam, K. Sabra, A. Erturk, 3D-printed gradient-index phononic crystal lens for underwater acoustic wave focusing. *Physical Review Applied* **13**, 064064 (2020).
54. E. Dong *et al.*, Bioinspired metagel with broadband tunable impedance matching. *Science advances* **6**, eabb3641 (2020).

55. E. Bok *et al.*, Metasurface for water-to-air sound transmission. *Physical review letters* **120**, 044302 (2018).

**Acknowledgments**

S. Q. wish to thank Zhen Dong, Walter Ho, Yuxiao Hou and Ho Yiu Mak for their assistance in the sample fabrication and shipping. S. Q. sincerely appreciates the helpful discussion with Min Yang on the integration scheme. N. G. wishes to thank Zhuoma Wang for her guidance in tensile test, William Wong for metal base fabrication and Shuang Nie for the Dynamic Mechanical Analysis (DMA) test. A.T. and B.M. thank H. Cahingt for the valuable technical assistance during the experiments.

**Funding:** P. S. acknowledges the support of RGC grant A-HKUST601/18 and AoE/P-502/20-3 for this work. J.P. G. and V. R.-G. acknowledge the support of the ANR-RGC METARoom project (ANR-18-CE08-0021) for this work.

**Author contributions:** S. Q. and N. G. contributed equally to this work. P. S. designed the research, conceived the idea, and supervised the project. S. Q. provided the theoretical framework. N. G. experimentally designed and characterized the composite materials. S. Q. and N. G. conducted the simulations and the sample fabrication. J.-P. G. and V. R.-G. supervised the sample shipment and the experiments in the water pool. B. M. and A. T. performed the experimental of the sample in the water pool. All the authors contributed to the data analysis and manuscript preparation.

**Competing interests:** Authors declare that they have no competing interests.

**Data and materials availability:** All data are available in the main text or the supplementary materials.

# Figures and Tables

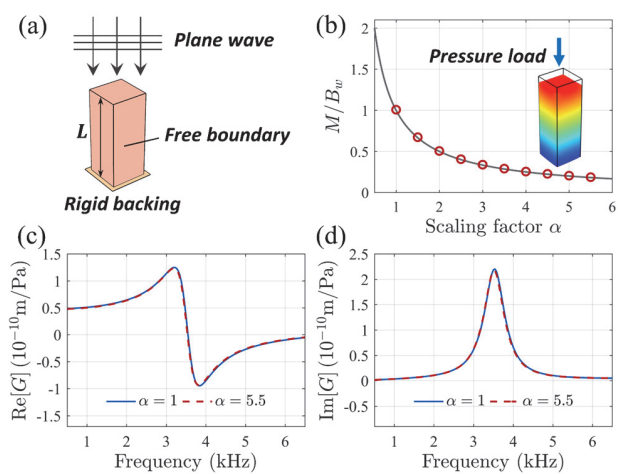

Fig. 1 The composite rod with tunable parameters. (a) The schematic of the single FP resonator made by a composite rod. (b) The effective modulus of the composite rod, as function of the scaling factor $\alpha$. The solid line denotes the Young's modulus $E_c = B_w/\alpha$ specified in the simulation, while the circles denote the effective longitudinal modulus, retrieved from the simulated output data. The inset figure shows the displacement field under the pressure load $\Delta p$. (c) The real part of the Green function, $G(\omega)$, with $\alpha = 1$ and 5.5. The data are retrieved from simulations. The complete overlap of the two curves shows that the Green function is independent of the scaling parameter $\alpha$. (d) The same for the imaginary part of the Green function.

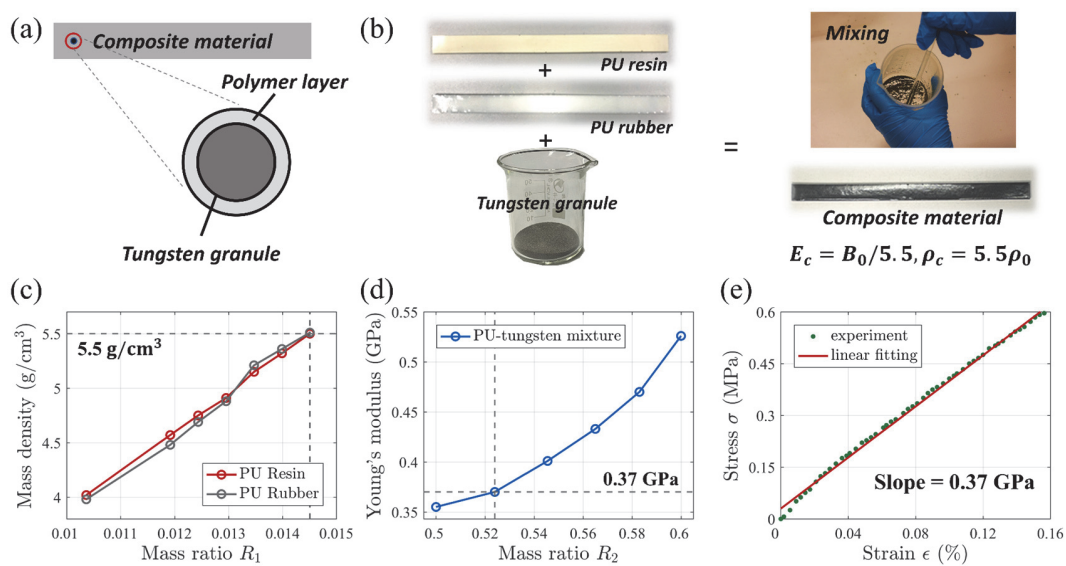

Fig. 2 The composite fabrication process and related tensile measurements for the composite with $\alpha = 5.5$ being the target. (a) A schematic illustration of the composite material's microstructure, comprising tungsten granules embedded in a polymer matrix. (b) By mixing PU resin, PU rubber and tungsten granules, we fabricate the composite material with the targeted properties, whose final stripe shape is used for the tensile test. (c) The relation between the mass density and the mass ratio $R_1 = M_{\mathrm{PU}}/M_{\mathrm{W}}$, for which $M_{\mathrm{PU}}$ can be contributed by either varying PU resin (red) or PU rubber (grey). (d) By fixing the mass ratio $R_1$, we change the mass ratio $R_2$ between the PU resin and PU rubber (i.e., $R_2 = M_{\mathrm{resin}}/M_{\mathrm{rubber}}$), in order to manipulate sample's Young's modulus. (e) The measured strain-stress curve for the final composite sample, showing its Young's modulus $E \cong 0.37 \times 10^9$ Pa, the target longitudinal modulus of the rod.

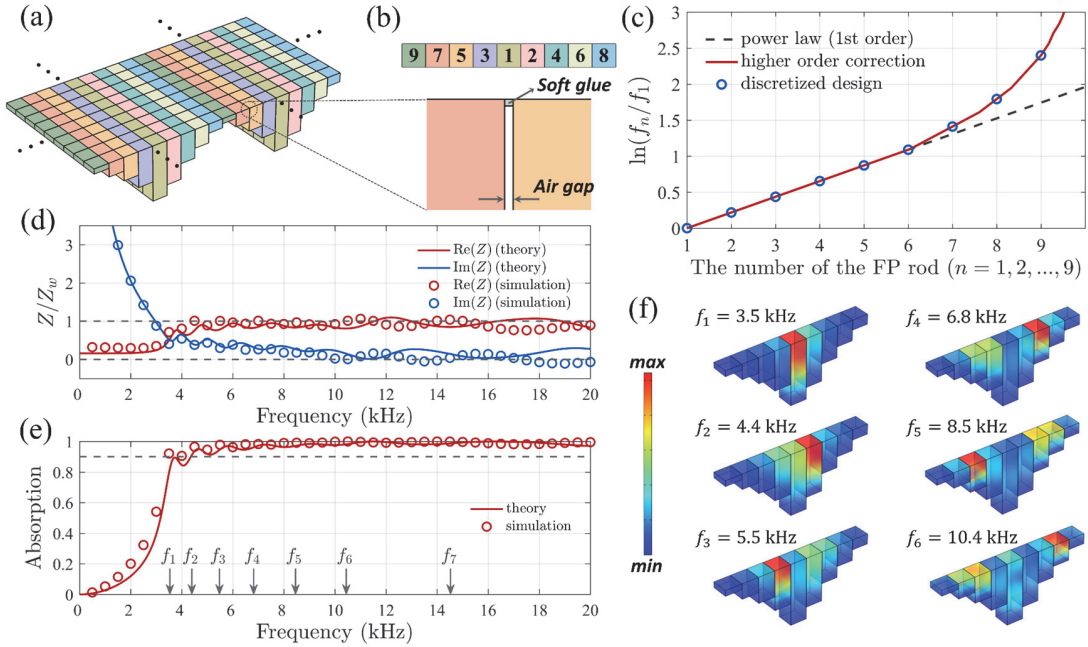

Fig. 3 The integrated FP resonators array, serving as the broadband underwater acoustic absorber. (a) The schematic of the resonator array. The dots indicate periodic repetition. (b) The nine FP resonator lengths, indicated by different colors together with an enlarged view of the air gap between two neighboring rods, with a thin layer of waterproof soft glue adhered on the top surface facing the water. (c) The design recipe of the resonance frequency distribution for realizing impedance matching. (d) The resulting real and imaginary parts of impedance, plotted as function of frequency. (e) The absorption spectrum and the first-order resonance frequencies are indicated by the grey arrows on the horizontal axis. Note that $f_8$ and $f_9$ are higher than 20 kHz. (f) The simulated velocity fields at different selected resonance frequencies.

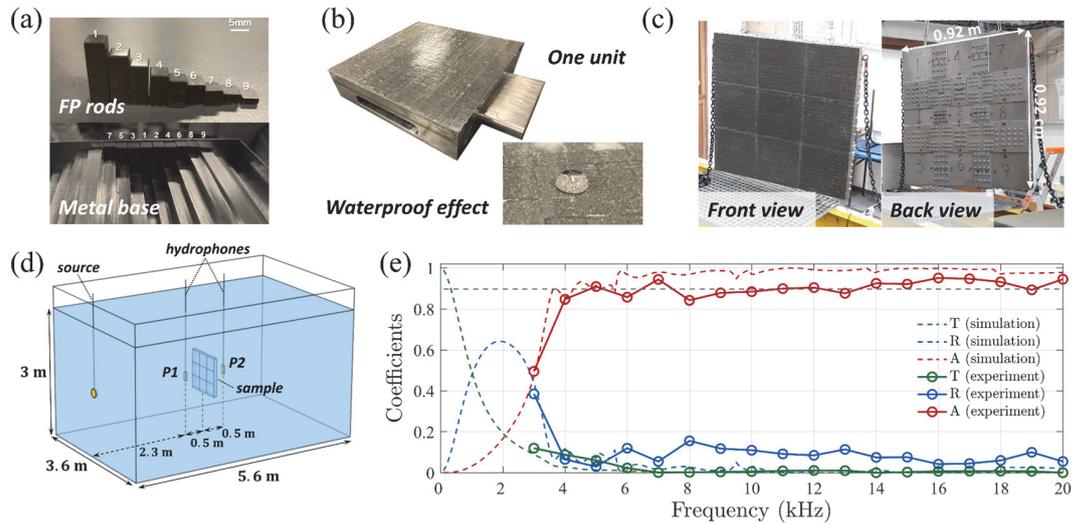

Fig. 4 Experimental realization of the underwater metamaterial absorber and the water pool measurement. (a) The fabricated composite rods, placed on the stepped stainless-steel base. (b) One sample unit and the waterproof effect on the top surface, with the adoption of the soft glue. Water is shown to form a droplet shape due to the hydrophobic effect of the waterproof soft glue. (c) The front view and back view of the assembled sample, comprising nine units. (d) The schematic of the water pool measurement configuration. (e) The measured data (circles) with the transmittance $T$, the reflectance $R$ and the absorption $A = 1 - T - R$. The results retrieved from simulation are also presented by dashed lines, for comparison. Here the simulations have taken into account the stainless-steel base.